\documentclass[twocolumn]{aastex631}
\usepackage{amsmath}
\usepackage{mathrsfs}
\usepackage{subfigure}
\usepackage{CJK}
\usepackage{multirow}

\shorttitle{Origins of $\chi_{\rm eff}$-$q$ correlation}
\shortauthors{Li et al.}

\graphicspath{{./}{figures/}}

\begin{document}
\begin{CJK*}{UTF8}{gbsn}

\title{Revealing the $\chi_{\rm eff}$-$q$ correlation among Coalescing Binary Black Holes and { Tentative} Evidence for AGN-driven Hierarchical Mergers}

\author[0000-0001-5087-9613]{Yin-Jie Li \textsuperscript{*} (李银杰)}
\affiliation{Key Laboratory of Dark Matter and Space Astronomy, Purple Mountain Observatory, Chinese Academy of Sciences, Nanjing 210023, People's Republic of China}
\email{ * Contributed equally.}

\author[0000-0001-9626-9319]{Yuan-Zhu Wang \textsuperscript{*} (王远瞩)}
\affiliation{Institute for Theoretical Physics and Cosmology, Zhejiang University of Technology, Hangzhou, 310032, People's Republic of China}

\author[0000-0001-9120-7733]{Shao-Peng Tang (唐少鹏)}
\affiliation{Key Laboratory of Dark Matter and Space Astronomy, Purple Mountain Observatory, Chinese Academy of Sciences, Nanjing 210023, People's Republic of China}

\author{Tong Chen (陈 彤)}
\affiliation{School of Physical Science and Technology, Inner Mongolia University, Hohhot 010021, People's Republic of China}

\author[0000-0002-8966-6911]{Yi-Zhong Fan (范一中)}
\affiliation{Key Laboratory of Dark Matter and Space Astronomy, Purple Mountain Observatory, Chinese Academy of Sciences, Nanjing 210023, People's Republic of China}
\affiliation{School of Astronomy and Space Science, University of Science and Technology of China, Hefei, Anhui 230026, People's Republic of China}
\email{The corresponding author: yzfan@pmo.ac.cn (Y.Z.F)}

\begin{abstract}
The origin of the correlation between the effective spins ($\chi_{\rm eff}$) and mass ratios ($q$) of LIGO-Virgo-KAGRA's binary black holes (BBHs) is still an open question. Motivated by the recent identification of two subpopulations of the BBHs, in this work we investigate the potential $\chi_{\rm eff}-q$ correlation for each subpopulation. Surprisingly, the $\chi_{\rm eff}$-$q$ correlation {either significantly weakens or disappears} for the low-mass subpopulation if we introduce a second $\chi_{\rm eff}$ distribution for the high-mass subpopulation, which likely originates from hierarchical mergers.  {This suggests that the $\chi_{\rm eff}$-$q$ correlation in the overall population can be explained by the superposition of two distinct subpopulations.}
{We find Bayesian evidence strongly favoring two separate $\chi_{\rm eff}$ distributions over a single mass-ratio-dependent distribution, with Bayes factors $\ln\mathcal{B}>4.2$.}
The first subpopulation has a narrow $\chi_{\rm eff}$ distribution peaking at $\sim0.05$, whose primary-mass function {showing a rapid decline beyond} $\sim 40M_{\odot}$, in agreement with first-generation BBHs. The second $\chi_{\rm eff}$ distribution is broad and peaks at $\mu_{\chi,2} \sim 0.4$, aligning with predictions for hierarchical mergers in active galactic nucleus (AGN) disks. 
{However, we cannot exclude negative $\chi_{\rm eff}$values in the second subpopulation, suggesting hierarchical mergers might occur both in AGN disks and stellar clusters. Furthermore, the inferred second $\chi_{\rm eff}$ distribution might alternatively arise from other formation channels, such as stable mass transfer or chemically homogeneous evolution, if not interpreted as hierarchical mergers.}

\end{abstract}


\section{Introduction}

The coalescing binary black holes (BBHs) will provide clues about their formation and evolutionary processes through the parameters of these systems \citep{2022PhR...955....1M}. In addition to characterizing the marginalized distributions of these parameters, it is also important to investigate the correlations among them \citep{2024PhRvD.109j3006H,2024arXiv241019145C}, including mass versus spin \citep[e.g.][]{2021ApJ...913L..19T,2022ApJ...941L..39W,2024PhRvL.133e1401L,2024ApJ...977...67L}, mass versus mass ratio \citep{2022ApJ...933L..14L}, spin versus mass ratio \citep{2021ApJ...922L...5C}, redshift versus spin \citep{2022ApJ...932L..19B}, and redshift versus mass versus spin \citep{2022ApJ...928..155T,2024ApJ...975...54G}.

\citet{2021ApJ...922L...5C}, for the first time, reported the anti-correlation between the effective spins ($\chi_{\rm eff}$) and mass ratios ($q$) of BBHs with data from GWTC-2 \citep{2019PhRvX...9c1040A, 2021PhRvX..11b1053A}. This anti-correlation was subsequently confirmed by \citet{2023PhRvX..13a1048A,2023ApJ...958...13A} with data from GWTC-3 \citep{2019PhRvX...9c1040A, 2021PhRvX..11b1053A,2024PhRvD.109b2001A,2023PhRvX..13d1039A}, {though the evidence is reduced with a more flexible model \citep{2024PhRvD.109j3006H}}. 
However, the origin of this anti-correlation is still in debate \citep{2021ApJ...922L...5C,2023PhRvX..13a1048A}.

Many simulations suggested that hierarchical mergers in the disks of active galactic nucleus (AGNs) could explain the observed $\chi_{\rm eff}$-$q$ correlation \citep[e.g.,][]{2022MNRAS.514.3886M,2023PhRvD.108h3033S,2024arXiv241110590C}.
The spin orientations and orbital angular momenta of binary black holes (BBHs) will be modulated by the disks, causing the mergers to favor positive effective spins. Additionally, the migration traps in AGN disks will produce mergers involving multiple generations of black holes (BHs) that have unequal masses and larger (positive) effective spins.
Previously, we found that the coalescing BHs can be divided into two subpopulations with significantly different spin-magnitude versus component-mass distributions, which are nicely consistent with first- and higher-generation BHs \citep{2024PhRvL.133e1401L}. 
{We also found that a fraction of the population that we interpreted as hierarchical mergers} have aligned spin tilts and asymmetric mass ratios, which may give rise to the $\chi_{\rm eff}$-$q$ anti-correlation.

However, other simulations suggested that certain formation channels of isolated binaries can produce highly spinning and unequal-mass BBHs, particularly the stable mass transfer formation scenario \citep{2024arXiv241115112B}.
Because the progenitors of unequal BBH systems in the stable mass transfer formation scenario are more likely to efficiently shrink their orbits during the second Roche-lobe overflow, which makes them easier to enter the tidal spin-up regime and later merge due to GW emission \citep{2024A&A...689A.305O}.
Therefore, it is possible that there may be a correlation between the effective spin and mass ratio in first-generation (or low-spin) BBHs, or among potential isolated evolution channels \citep{2022ApJ...941L..39W,2023arXiv230401288G,2024ApJ...977...67L}.

In this study, we delve into the question of whether the anti-correlation between $\chi_{\rm eff}$-$q$ in BBHs originates from a superposition of various formation channels / subpopulations \citep{2024PhRvL.133e1401L},  if it emerges from the evolutionary processes of a single population with a common formation channel \citep[e.g.,][]{2024arXiv241115112B}, or alternatively, both effects have contributed to the observed correlation.
Additionally, we investigate the $\chi_{\rm eff}$ distribution in the high-mass range that {might} corresponds to the hierarchical mergers, in order to find out whether these events originate from the AGN disks, or alternatively from the star clusters which exhibits a symmetric $\chi_{\rm eff}$ distribution \citep{2024ApJ...966L..16P,2025PhRvL.134a1401A}.

This work is organized as follows: In Section~\ref{sec:method} and Section~\ref{sec:results}, we introduce the methods and the results. In Section~\ref{sec:diss}, we present the conclusions and engage in discussions.

\section{Methods}\label{sec:method}

We use hierarchical Bayesian inference to measure the hyperparameters of population model; see Appendix~\ref{app:meth} for details. 
Following \citet{2023PhRvX..13a1048A}, we adopted 69 BBH events with false alarm rates (FAR) $<1 {\rm yr}^{-1}$ in GWTC-3 for analysis. The posterior samples for each BBH event are obtained from \href{https://zenodo.org/doi/10.5281/zenodo.5546662}{events-zenodo}, and the `C01:Mixed' samples are adopted.

We first use the same $\chi_{\rm eff}-q$ distribution model as that used in \citet{2021ApJ...922L...5C} to fit data of GWTC-3 for comparison.
Specifically, the mean and width of the $\chi_{\rm eff}$ distribution change linearly with $q$ (referred to as the Base model); see Appendix~\ref{model_xq} for the detailed formula. The rate evolution model is the MD model \citep{2014ARA&A..52..415M}, as defined in the Appendix~\ref{app:zmodel}. 
{We use both the non-/semi-parametric formula \textsc{PowerLaw+Spline} \citep[PS;][]{2022ApJ...924..101E,2023PhRvX..13a1048A} and the popular parametric formula \textsc{PowerLaw+Peak} \citep[PP;][]{2018ApJ...856..173T,2021ApJ...913L...7A} to model the primary-mass distribution, see Appendix~\ref{m_model} for the details of mass functions.} 
With the PP model, we find an anti-correlation between $\chi_{\rm eff}$ and $q$ distribution (i.e., $a<0$) at 99.2\% credible level. However, the credibility decreases to 93\% with the more flexible PS model, 
while the overall tendency remains unchanged, as shown in Figure~\ref{fig:compare_trace}. 

Inspired by the investigation of spin-magnitude versus component-mass distribution \citep{2024PhRvL.133e1401L}, we introduce another $\chi_{\rm eff}$-distribution to capture the secondary subpopulation of BBHs that is consistent with hierarchical mergers. Two kind of models are applied, one is the Mixture model on primary-mass versus effective-spin distribution,
\begin{equation}
\begin{aligned}
\pi_{\rm mix}&(m_1,m_2,\chi_{\rm eff} |{\bf \Lambda}) = P(m_2|m_1;{\bf \Lambda})\\
&[\pi_1(m_1,\chi_{\rm eff} |{\bf \Lambda})(1-r_2)+\pi_2(m_1,\chi_{\rm eff}|{\bf \Lambda})r_2],
\end{aligned}
\end{equation}
with
\begin{equation}
\begin{aligned}
\pi_1&(m_1,\chi_{\rm eff} |{\bf \Lambda})=P_{1,m_1}(m_1|{\bf \Lambda})P(\chi_{\rm eff}|q;\mu_{\chi,0},\sigma_{\chi,0},a,b)
\end{aligned}
\end{equation}
and 
\begin{equation}
\begin{aligned}
\pi_2&(m_1,\chi_{\rm eff}|{\bf \Lambda})=\mathcal{PL}(m_1|-\alpha_2,m_{\rm min,2},m_{\rm max,2})\\
&\mathcal{G}_{[-1,1]}(\chi_{\rm eff}|\mu_{\chi,2}, \sigma_{\chi,2})
\end{aligned}
\end{equation}
where $r_2$ is the mixed fraction of the second subpopulation, and $\mathcal{G}_{[-1,1]}$ is the truncated Gaussian distribution. $P_{m_1}$ and $\mathcal{PL}$ are the primary-mass functions for the two subpopulation, {and we adopt $\mathcal{PS}$ and $\mathcal{PP}$ for $P_{m_1}$, respectively.} $P(m_2|m_1;{\bf \Lambda})$ is the normalized secondary-mass distribution conditioned on $m_1$, $P(\chi_{\rm eff}|q;\mu_{\chi,0},\sigma_{\chi,0},a,b)$ is the Base model that encodes the $\chi_{\rm eff}-q$ correlation \citep{2021ApJ...922L...5C}, see Appendix~\ref{app:model} for the detailed formulas of the models. { Note that we do not apply either the low-mass tapering function or the perturbation function to the power-law mass function for the second subpopulation, to avoid overly complex (bloated) models. Actually, with currently available data, the results are rather similar when the low-mass tapering and perturbation functions are applied, see Appendix~\ref{app:results}.
}
 
The other model has a more concise formula, i.e., the two $\chi_{\rm eff}$-distributions are modulated by a transition function of the primary mass \citep[see also][for similar constructions]{2022ApJ...941L..39W,2024PhRvL.133e1401L,2024ApJ...975...54G}, hereafter Transition model,
\begin{equation}
\begin{aligned}
\pi_{\rm tran}&(m_1,m_2,\chi_{\rm eff} |{\bf \Lambda}) = P_{m_1}(m_1|{\bf \Lambda})P(m_2|m_1;{\bf \Lambda})\\
&P_{\rm tran}(\chi_{\rm eff} |q,m_1;{\bf \Lambda}),
\end{aligned}
\end{equation}
with
 \begin{equation}
\begin{aligned}
P_{\rm tran}(\chi_{\rm eff} &|q,m_1;{\bf \Lambda})=\\
&P(\chi_{\rm eff}|q;\mu_{\chi,0},\sigma_{\chi,0},a,b)\frac{1}{1+e^{(m_1-m_{\rm t})/\delta_{\rm t}}} +\\
&\mathcal{G}_{[-1,1]}(\chi_{\rm eff}|\mu_{\chi,2}, \sigma_{\chi,2})  \frac{1}{1+e^{(m_{\rm t}-m_1)/\delta_{\rm t}}} ,
\end{aligned}
\end{equation}
where $m_{\rm t}$ and $\delta_{\rm t}$ are the location and rapidness of the transition. For the BBHs with primary mass below / above $m_{\rm t}$, the $\chi_{\rm eff}$ distribution is dominated by the Base model / second Gaussian distribution.

\section{Results}\label{sec:results}

\begin{figure*}[!htb]
    \centering
    \subfigure{\includegraphics[width=0.72\hsize]{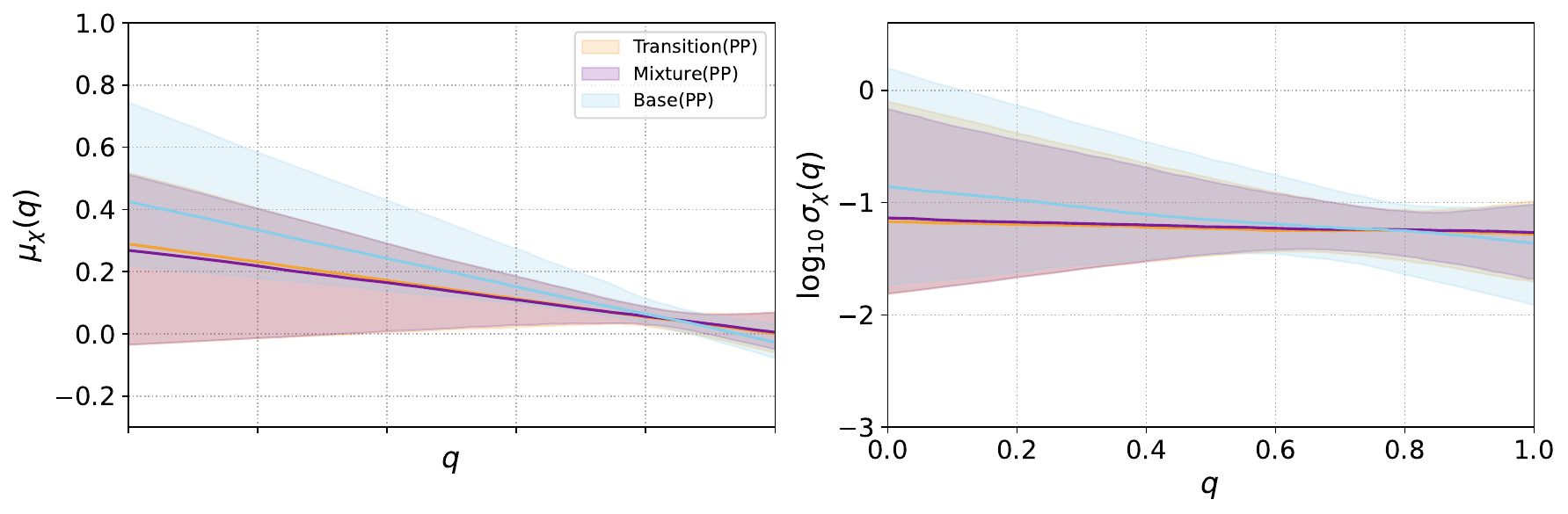}\label{fig: sub_figure1}}
    \subfigure{\includegraphics[width=0.22\hsize]{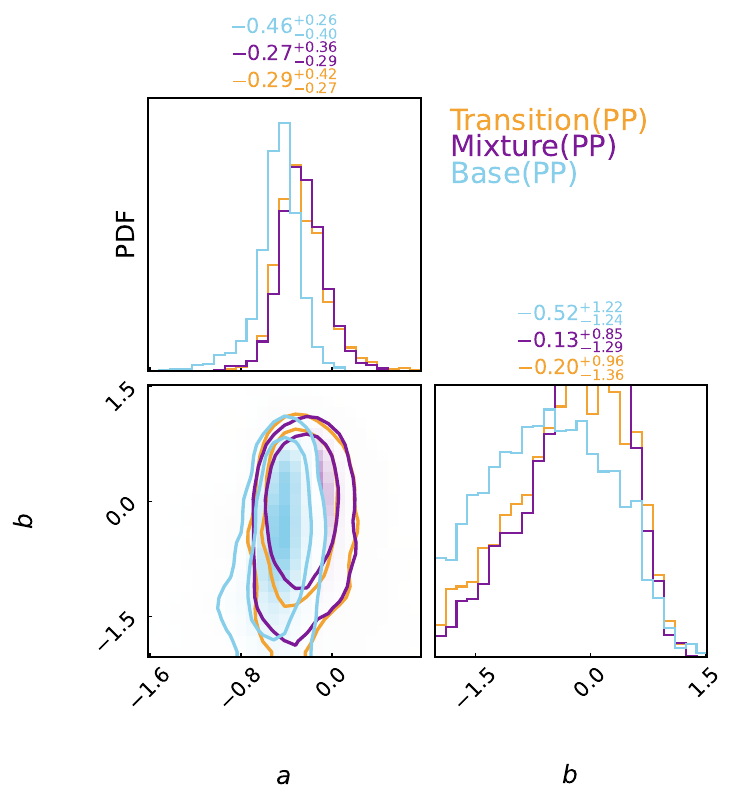}\label{fig: sub_figure3}}
    \subfigure{\includegraphics[width=0.72\hsize]{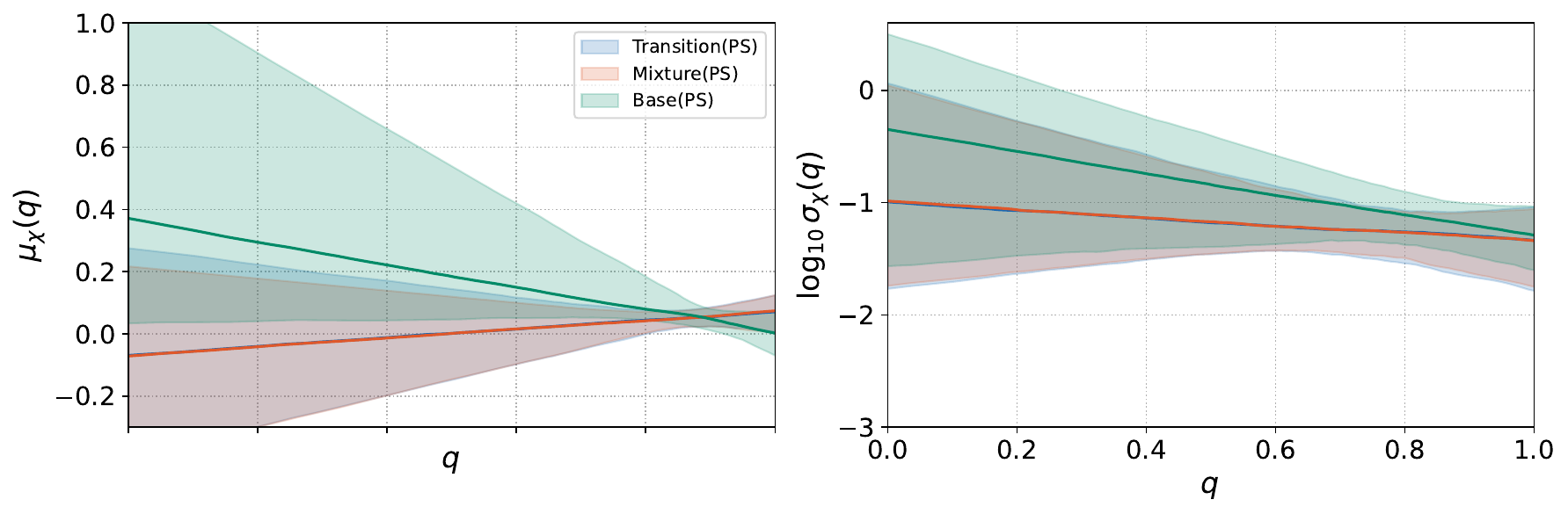}\label{fig: sub_figure2}}
    \subfigure{\includegraphics[width=0.22\hsize]{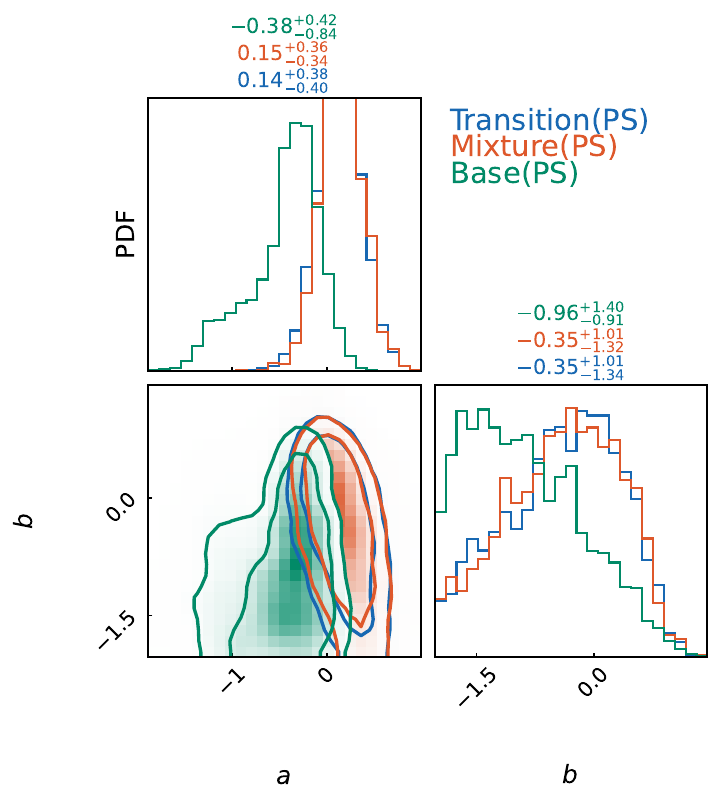}\label{fig: sub_figure4}}
    \caption{Left\&Mid: Constraints on the mean $\mu_{\chi}(q)$ and standard deviation $\sigma_{\chi}(q)$ of the $\chi_{\rm eff}$ distribution, as a function of BBH mass ratio $q$. The solid curves are the medians and the colored bands are the 90\% credible intervals.
Right: Posteriors of the hyperparameters describing the $\chi_{\rm eff}-q$ correlation. The contours mark the central 50\% and 90\% posterior credible regions, the values represent the median and 90\% credible intervals.}
    \label{fig:compare_trace}
\end{figure*}

In Table~\ref{tab:results}, we summarize the Bayes factors of the novel models in this work compared to the {Base (PS) and Base (PP) models, respectively. For both PS and PP cases, the Transition model and Mixture model are favored than the Base models.} Furthermore, the Transition model and Mixture model without the $\chi_{\rm eff}$-$q$ correlation (i.e., $a=0,b=0$) are even more favored, {with $\ln\mathcal{B}>4.2$. }
Figure~\ref{fig:compare_trace}(right panels) shows the posteriors of the hyperparameters describing the $\chi_{\rm eff}$ and $q$ correlation. In each case, the parameters of the Transition model and Mixture model are highly consistent with each other. For the PS case, the slope parameters $a$ and $b$ are consistent with zero. {For the PP case, the $\chi_{\rm eff}$-$q$ correlation still remains, but the credibility of $a<0$ decreases to $\sim90\%$ comparing to the Base (PP) model. It is consistent in both cases that the anti-correlation reduces when we introduce a second subpopulation of $\chi_{\rm eff}$ distribution, which means that the second subpopulation has contributed to the $\chi_{\rm eff}$-$q$ correlation in the whole population. Specifically, the anti-correlation of the entire BBH population mainly results from the superposition of two subpopulations. However, whether there is $\chi_{\rm eff}$-$q$ correlation in the first subpopulation remains uncertain. Additionally, we find there is no evidence for the $\chi_{\rm eff}$-$q$ correlation in the second subpopulation with currently available data, see Appendix~\ref{app:2ab}.}
In subsequent analysis, we adopt the results inferred with $a=0,b=0$, since the posterior distributions of parameters are broadly consistent with the results inferred with variable $a$ and $b$, see Appendix~\ref{app:results}.

\begin{table}[htpb]
\centering
\caption{Model comparison}\label{tab:results}
\begin{tabular}{lcc}
\hline
\hline
Model     &  $\ln{\mathcal{B}}_{\rm PS}^{X}$  \\
\hline
Base(PS)   & 0  \\
Base(PS) with $a=0$,$b=0$  & 0  \\
Transition(PS) & 4.9 \\
Transition(PS) with $a=0$,$b=0$ & 6.7 \\
Mixture(PS) & 3.8 \\
Mixture(PS) with $a=0$,$b=0$ & 6.1 \\
Base(PP) & -5.9 \\
\hline
     &  $\ln{\mathcal{B}}_{\rm PP}^{X}$  \\
\hline
Base(PP) & 0 \\
Base(PP) with $a=0$,$b=0$  & -1.2 \\
Transition(PP) & 3.2 \\
Transition(PP) with $a=0$,$b=0$ & 4.9 \\
Mixture(PP)  & 2.7 \\
Mixture(PP) with $a=0$,$b=0$ & 4.2 \\
\hline
\hline
\end{tabular}
\\
\begin{tabular}{l}
Note: All the log Bayes factors in the top / bottom are\\
 relative to the Base (PS) / (PP) model that encode the\\
  $\chi_{\rm eff}$-$q$ correlation and accompanied with the PS / PP\\
   mass fucntion. 
\end{tabular}
\end{table} 

\begin{figure}
	\centering  
\includegraphics[width=0.96\linewidth]{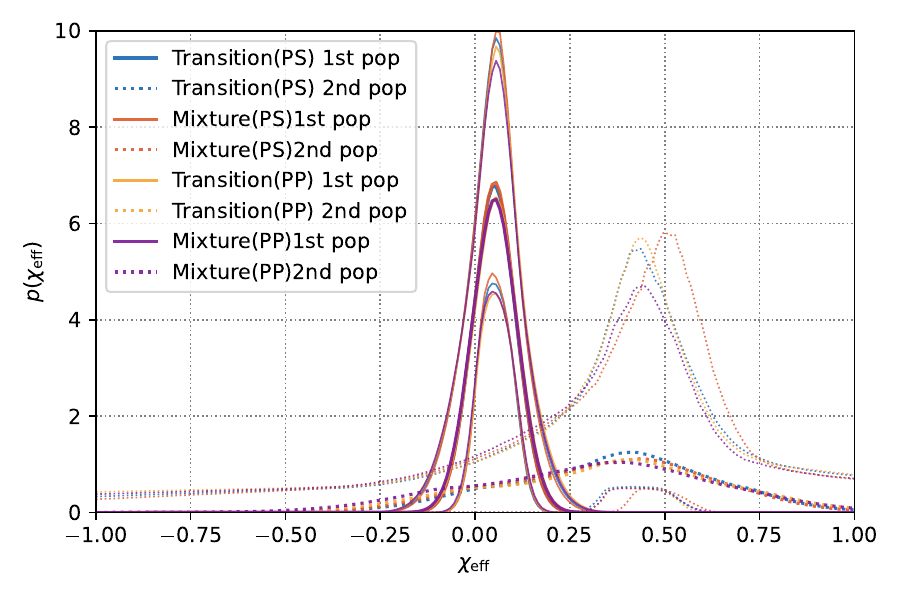}
\caption{Effective-spin distributions of the two subpopulation inferred by the Transition and Mixture models with $a=0$,$b=0$. The lines represent the mean values and 90\% credible intervals.}
\label{fig:chi_dist}
\end{figure}

We find that the second subpopulation shows a significantly different $\chi_{\rm eff}$ distribution compared to the first subpopulation, see Figure~\ref{fig:chi_dist}. 
The divergence point of the Transition model in the primary-mass function is $m_{\rm t}=49^{+13}_{-9}M_{\odot}$ \footnote{In this work, when quoting $X^{+Y}_{-Z}$, X means the posterior median and $[X-Y,Y+Z]$ means the central 90\% credible interval.}, with a transition scale of {$\delta_{\rm t}=6^{+3}_{-4}M_{\odot}$ ($\delta_{\rm t}=6^{+4}_{-4}M_{\odot}$) for the PS (PP) case. Note that $m_{\rm t}$ and $\delta_{\rm t}$ are degenerated with each other, and $\delta_{\rm t}$ is only loosely constrained, see Figure~\ref{fig:mt_corner}.}

As for the Mixture model, 
{though the maximum mass of the first subpopulation is not well constrained, the masses of 99.5th and 99.75th, and 99.9th percentiles are better measured. Specifically, the $m_{99.9\%}$ of the first subpopulation is significantly different from that of the entire population, see Appendix~\ref{app:results}.}
The minimum mass of the second subpopulation is $\sim 30 M_{\odot}$. These results are consistent with our previous analysis using spin magnitudes of BBHs \citep{2022ApJ...941L..39W,2024PhRvL.133e1401L}, see Appendix~\ref{app:results} for the distributions of other related parameters. The $\chi_{\rm eff}$ distributions of both subpopulations are not symmetric with respect to zero, indicating the presence of formation channels other than star clusters. 

{We find that $\mu_{\chi,2}>0$ at 98\% (95\%) credible level, and a symmetric $\chi_{\rm eff}$ distribution with respect to zero for the second population (i.e.,$\mu_{\chi,2}=0$) is disfavored by a Bayes factor of $\ln\mathcal{B}=1.4$ ($\ln\mathcal{B}=1.5$) for the PS (PP) case, which indicates that the hierarchical mergers cannot be produced (solely) by star clusters. The second $\chi_{\rm eff}$ distribution is broad and peaks at $\mu_{\rm eff,2}=0.40^{+0.32}_{-0.30}$ ($\mu_{\rm eff,2}=0.39^{+0.39}_{-0.37}$) for the PS (PP) case, which is consistent with the hierarchical mergers in the AGN disks \citep{2019PhRvL.123r1101Y}. However, some other formation channels, like chemically homogeneous evolution \citep{2016MNRAS.458.2634M} and stable mass transfer \citep{2021A&A...647A.153B} can not be ruled out  based solely on the effective spin distribution \citep{2021ApJ...910..152Z}.} 
{We have also inferred using two more flexible model for the second $\chi_{\rm eff}$ distribution, one with variable edges for the truncated Gaussian, the other is spline function \citep{2023PhRvD.108j3009G}, see Appendix~\ref{app:asym}. We find that the second $\chi_{\rm eff}$ distribution is not symmetric about zero, but having preference for positive values.  Additionally, the negative values for the second $\chi_{\rm eff}$ distributions can not be ruled out. 
} 

\section{Conclusions and Discussion}\label{sec:diss}

In this work, we investigate the origins of $\chi_{\rm eff}$-$q$ correlation in the BBHs \citep{2021ApJ...922L...5C,2023PhRvX..13a1048A} with data of GWTC-3, 
using the dedicated models that introduce a second $\chi_{\rm eff}$ distribution for the second (high-spin) subpopulation. 

{In both PS and PP cases, the $\chi_{\rm eff}$-$q$ correlation significantly reduces when we include a second subpopulation for the $\chi_{\rm eff}$ distribution. Additionally, there is no evidence for a $\chi_{\rm eff}-q$ correlation in the second subpopulation, see Appendix~\ref{app:2ab}. Therefore we can conclude that the $\chi_{\rm eff}$-$q$ anti-correlation of the entire BBH population mainly results from {\it the superposition of two subpopulations}. However, whether there is $\chi_{\rm eff}$-$q$ correlation in the first subpopulation remains uncertain. Because for the PS case, $a$ is consistent with zero, while for the PP case, we still find $a<0$ at $\sim90\%$ credible level, see Figure~\ref{fig:compare_trace}.
} 

{The stable mass transfer formation channel is expected to produce BBHs with $\chi_{\rm eff}-q$ anti-correlation \citep{2024A&A...689A.305O,2024arXiv241115112B}, while BBHs from the common envelop channel may exhibit an opposite correlation \citep{2020A&A...635A..97B,2021A&A...647A.153B}. Therefore, it is possible that there is a $\chi_{\rm eff}$-$q$ correlation in the first subpopulation, which is consistent with the first-generation BBHs. Note that the first subpopulation may consist not only the isolated formation channels but also the dynamical formation channels \citep{2024arXiv240403166R,2024ApJ...977...67L}. When the detections are enriched, more subpopulations may be recognized and the correlations of their parameters may also be measured.}

{
The $\chi_{\rm eff}$ distribution of the second subpopulation (shown in Figure~\ref{fig:chi_dist}), is consistent with the hierarchical mergers in gas-rich environments, such as AGN disks \citep{2019PhRvL.123r1101Y}.
Some other formation channels may also produce $\chi_{\rm eff}$ distributions similar to that of the second subpopulation inferred with our models \citep[see][]{2021ApJ...910..152Z}, such as chemically homogeneous evolution \citep{2016MNRAS.458.2634M} and stable mass transfer \citep{2021A&A...647A.153B}. However, the second subpopulation found in this work is actually consistent with the population of hierarchical mergers found in our previous works \citep{2024PhRvL.133e1401L}, where we have provided smoking-gun evidence, specifically a spin-magnitude distribution of approximately 0.7 \citep{2021NatAs...5..749G}. Therefore, hereafter we simply interpolate the second subpopulation as hierarchical mergers.}
\citet{2025PhRvL.134a1401A} attributed this subpopulation to hierarchical mergers in star clusters, as they found that the $\chi_{\rm eff}$ distribution is consistent with a uniform distribution in the range of approximately (-0.5, 0.5). 
{
To check out whether such difference is caused by the systematic bias in our analysis, we perform end-to-end mock data studies including the parameter estimation and hierarchical analysis of mock signals, see Appendix~\ref{app:mock}. We find that the two subpopulations in the $\chi_{\rm eff}$ distribution can be successfully recognized. Additionally, we repeat the simulation for many times, the recoveries of the second $\chi_{\rm eff}$ distributions are always broadly consistent with the injections. When injected distribution is symmetric (asymmetric), then the recovered distributions are almost symmetric (asymmetric). Thus, the identified asymmetry in the second subpopulation’s $\chi_{\rm eff}$ distribution in our study is attributed to the data itself rather than to methodological biases.
With currently available data, we find the symmetric $\chi_{\rm eff}$ distribution of the second subpopulation ($\mu_{\chi,2}$) is less favored by $\ln\mathcal{B}\sim1.5$. A more conclusive evidence may be found when the data from the fourth observing run\footnote{See https://gracedb.ligo.org/superevents/public/O4/ for the catalog of detection candidates.} are released.} 

{Many simulations \citep[e.g.,][]{2019PhRvL.123r1101Y,2023PhRvD.108h3033S,2024arXiv241110590C} show that the hierarchical mergers in disk-like environments can produce a $\chi_{\rm eff}$ distribution peaking at $\sim0.4$ that is consistent with our results. Additionally, AGN-driven mergers (including first generations) will exhibit a $\chi_{\rm eff}$-$q$ correlation that is found in the entire BBH population \citep{2023PhRvD.108h3033S,2024arXiv241110590C}. Therefore, it is very likely that AGN-disk channel has contributed to the BBHs detected by LVKC \citep{2019PhRvX...9c1040A, 2021PhRvX..11b1053A,2024PhRvD.109b2001A,2023PhRvX..13d1039A}.
Some works also suggest that the BBHs with primary masses at $\sim 35 M_{\odot}$ peak are consistent with dynamical formation channels in star clusters \citep{2024ApJ...977...67L,2024arXiv240403166R}, as predicted by \cite{2023MNRAS.522..466A}. Therefore, the star clusters have also contributed to the hierarchical mergers, if the remnants of the previous mergers are sufficiently retained in the environments \citep{2021NatAs...5..749G,2022ApJ...935L..20Z,2024arXiv241109195L}. Our further analysis with a more flexible models also show that the negative values for the second $\chi_{\rm eff}$ distributions can not be ruled out, see Appendix~\ref{app:asym}. 
}

However, there are only $\sim 11$ hierarchical mergers (high-spin events) in GWTC-3 \citep{2024PhRvL.133e1401L}, making it difficult to determine the mixture fractions of AGN-disk like and star-cluster formation channels. We will address such issue when the GW data are significantly enriched.
The fourth observing run (O4) of the LIGO-Virgo-KAGRA GW detectors is currently underway, and the number of detections is rapidly increasing (see \url{https://gracedb.ligo.org/latest/}). At the end of O4, more than four times as many events are expected to be observed compared to O3 \citep{2024arXiv241019145C}. With the enriched data more subpopulations / formation channels of BBHs may be identified \citep{2021ApJ...910..152Z}, and the mixture fractions and the parameter correlations of subpopulations can be better determined \citep{2022ApJ...933L..14L,2024ApJ...977...67L,2024ApJ...975...54G,2024PhRvD.109j3006H,2024arXiv241019145C}. 
     
%%%%%%%%%%%%%%%%%%%%%%%%%%%%%%%%%%%%%%%%%%%%%%%%%%%%%%%%%%%%%
% Acknowledgments
%%%%%%%%%%%%%%%%%%%%%%%%%%%%%%%%%%%%%%%%%%%%%%%%%%%%%%%%%%%%%

\begin{acknowledgments}
This work is supported by the National Natural Science Foundation of China (No. 12233011), the General Fund (No. 2024M753495) of the China Postdoctoral Science Foundation, and the Priority Research Program of the Chinese Academy of Sciences (No. XDB0550400). Y-Z Wang is supported by  the National Natural Science Foundation of China (No. 12203101) and S-P Tang is supported by the National Natural Science Foundation of China (No. 12303056), the General Fund (No. 2023M733736) and the Postdoctoral Fellowship Program (GZB20230839) of the China Postdoctoral Science Foundation. This research has made use of data and software obtained from the Gravitational Wave Open Science Center (https://www.gw-openscience.org), a service of LIGO Laboratory, the LIGO Scientific Collaboration and the Virgo Collaboration. LIGO is funded by the U.S. National Science Foundation. Virgo is funded by the French Centre National de Recherche Scientifique (CNRS), the Italian Istituto Nazionale della Fisica Nucleare (INFN) and the Dutch Nikhef, with contributions by Polish and Hungarian institutes.
\end{acknowledgments}

\vspace{5mm}

\software{Bilby \citep[version 1.1.4, ascl:1901.011, \url{https://git.ligo.org/lscsoft/bilby/}]{2019ascl.soft01011A},
          %Dynesty \citep[version 1.0.1, \url{https://github.com/joshspeagle/dynesty}]{2020MNRAS.493.3132S},
          PyMultiNest \citep[version 2.11, ascl:1606.005, \url{https://github.com/JohannesBuchner/PyMultiNest}]{2016ascl.soft06005B}.
          %Icarogw \citep[\url{https://git.ligo.org/cbc-cosmo/icarogw}]{2021PhRvD.104f2009M,2023arXiv230517973M}.
          %Nessai \citep[\url{https://nessai.readthedocs.io/en/latest/}]{Williams:2021qyt,Williams:2023ppp,nessai}
          }
                      
%%%%%%%%%%%%%%%%%%%%%%%%%%%%%%%%%%%%%%%%%%%%%%%%%%%%%%%%%%%%%
% Appendix
%%%%%%%%%%%%%%%%%%%%%%%%%%%%%%%%%%%%%%%%%%%%%%%%%%%%%%%%%%%%%

\appendix
\section{Hierarchical Bayesian Inference}\label{app:meth}

We perform hierarchical Bayesian inference to infer the hyperparameters $\Lambda$ describing population models $\pi(\theta|\Lambda)$. Following the framework described in \citep{2019MNRAS.486.1086M,2021ApJ...913L...7A,2023PhRvX..13a1048A}, given $\Lambda$, the likelihood of the GW data $\{d\}$ from $N_{\rm det}$ detections can be expressed as, 
\begin{equation}
\mathcal{L}(\{d\}|\Lambda)\propto N^{N_{\rm det}}e^{-N_{\rm exp}} \prod_{i}^{N_{\rm det}}{\int{\pi(\theta_i|\Lambda)\mathcal{L}(d_i|\theta_i)d\theta_i}},
\end{equation}
where $N$ is the total number of mergers in the surveyed time-space volume, which is related to the merger rate density over cosmic history $N=\int{R(z|\Lambda)\frac{dV_{\rm c}}{dz}\frac{T_{\rm obs}}{1+z}dz}$.
$N_{\rm exp}$ is the expected number of detections, which is related to the detection probability $P({\rm det}|\theta)$, i.e., $N_{\rm exp}=N\int{P({\rm det}|\theta)\pi(\theta|\Lambda)d\theta}$. This term can be calculated using a Monte Carlo integral over the referred injection\footnote{Adopted from https://zenodo.org/doi/10.5281/zenodo.5636815.}, see Appendix of \cite{2021ApJ...913L...7A} for details. 
$\mathcal{L}(d_i|\theta_i)$ is the likelihood of the $i$-th event, which can be evaluated using the posterior samples \citep[see][for detailed illustration]{2021ApJ...913L...7A}.

Following \citet{2023PhRvX..13a1048A,2023MNRAS.526.3495T}, we define the effective number of samples for the $i$-th event in the Monte Carlo integral as $N_{{\rm eff},i}=\frac{[\sum_{j}{w_{i,j}]^2}}{\sum_j{w_{i,j}^2}}$, where $w_{i,j}$ is the weight of $j$-th sample in $i$-th event. We constrain $N_{{\rm eff},i} > 10$ to ensure accurate evaluation of likelihood, which is sufficiently high given the sample size of GWTC-3 \citep{2022arXiv220400461E}. Additionally, we constrain the effective number of found injections remaining after population reweighting as $N_{\rm eff,sel}> 4 N_{\rm det}$, to ensure an accurate estimation of $N_{\rm exp}$ \citep{2019RNAAS...3...66F,2021ApJ...913L...7A}.

\section{Population models}\label{app:model}

\subsection{Mass function}\label{m_model}
The parametric primary-mass function is the popular \textsc{PowerLaw+Peak} model \citep{2021ApJ...913L...7A} , which reads,
\begin{equation}
\begin{aligned}
\mathcal{PP}&(m_1|{\bf \Lambda}) \propto \\ 
&[\mathcal{PL}(m_1|-\alpha,m_{\rm min},m_{\rm max})(1-\lambda_{\rm peak})+\\
&\mathcal{G}_{[m_{\rm min},m_{\rm max}]}(m_1|\mu_{\rm m},\sigma_{\rm m})\lambda_{\rm peak}]
\times \mathcal{S}(m_1|\delta_{\rm m},m_{\rm min}),
\end{aligned}
\end{equation}
where $\mathcal{PL}(m_1|-\alpha,m_{\rm min},m_{\rm max})$ is the Power-law distribution with slope index of $-\alpha$ truncated on ($m_{\rm min}$, $m_{\rm max}$). $\mathcal{G}$ is the Gaussian distribution with mean $\mu_{\rm m}$ and width $\sigma_{\rm m}$ truncated on ($m_{\rm min}$, $m_{\rm max}$), $\mathcal{S}(m_1|\delta_{\rm m},m_{\rm min})$ is the smooth function with smooth scale of $\delta_{\rm m}$, impacting on the low edge $m_{\rm min}$.

In order to reduce the bias that may be brought by the mis-specification of parametric formulas, we use a semi-/non-parametric mass function\citep{2022ApJ...924..101E} for the main analysis,
\begin{equation}
\begin{aligned}
\mathcal{PS} & (m_1|{\bf \Lambda}) \propto \mathcal{PL}(m_1| -\alpha, m_{\rm min}, m_{\rm max})\\
&\times \mathcal{S}(m_1|\delta_{\rm m},m_{\rm min}) e^{f(m|\{x_i\},\{f_{i}\})},
\end{aligned}
\end{equation}
where $f(m|\{x_i\},\{f_{i}\})$ is the cubic-spline perturbation function interpolated between the knots ($x_i,f_i$) placed in the mass range. Here we adopted 12 knots $\{x_i\}_{i=0}^{12}$ linearly distributed in log space of (6, 80)$M_{\odot}$, and restrict the perturbation to zero at the minimum and maximum knots.

The secondary-mass function is conditioned on the primary mass \citep{2021ApJ...913L...7A}, 
\begin{equation}
P(m_2|m_1;{\bf \Lambda})\propto\mathcal{PL}(m_2|\beta,m_{\rm min},m_1)\mathcal{S}(m_2|\delta_{\rm m},m_{\rm min}).
\end{equation}

\subsection{Effective-spin  distribution model} \label{model_xq}
We follow \citet{2021ApJ...922L...5C} to construct a mass-ratio-dependent model for effective-spin distribution, i.e., the Base model,
\begin{equation}
\begin{aligned}
P&(\chi_{\rm eff}|q;\mu_{\chi,0},\sigma_{\chi,0},a,b)=\\
&\mathcal{G}_{[-1,1]}(\mu_{\chi}(q;\mu_{\chi,0},a), \sigma_{\chi}(q;\sigma_{\chi,0},b)),
\end{aligned}
\end{equation}
with 
\begin{equation}
\begin{aligned}
\mu_{\chi}(q;\mu_{\chi,0},a)=\mu_{\chi,0}+a(q-0.5),\\
\log\sigma_{\chi}(q;\sigma_{\chi,0},b)=\log\sigma_{\chi,0}+b(q-0.5),
\end{aligned}
\end{equation}
where $\mathcal{G}_{[-1,1]}$ is the Gaussian truncated on $[-1,1]$. Note that the linear functions in \citet{2023PhRvX..13a1048A} is slightly different, where the authors let $\mu_{\chi}(q=1)=\mu_{\chi,0}$ and $\log\sigma_{\chi}(q=1)=\log\sigma_{\chi,0}$.

\subsection{Rate evolution model}\label{app:zmodel}

The merger rate density as a function of redshift reads \citep[MD model][]{2014ARA&A..52..415M},
\begin{equation}\label{eq:zmodel}
\begin{aligned}
&R(z|\gamma,\kappa,z_{\rm p})=R_0\times\frac{[(1+z_{\rm p})^{(\gamma+\kappa)}+1](1+z)^{\gamma}}{(1+z)^{(\gamma+\kappa)}+(1+z_{\rm p})^{(\gamma+\kappa)}},
\end{aligned}
\end{equation}
where $R_0$ is the local merger rate density. Note the injection campaign only provides mock events with $z<1.9$, so we normalize the redshift distribution $P(z|\gamma,\kappa,z_{\rm p})$ within $(0,1.9)$, when calculating likelihood.

\begin{table*}[htpb]
\centering
\caption{Summary of model parameters.}\label{tab:prior}
\begin{tabular}{lcccc}
\hline
\hline
Parameter     &  Description & Prior \\
\hline
$m_{\rm min}[M_{\odot}]$   & The minimum mass & $U(2,10)$  \\
$m_{\rm max}[M_{\odot}]$   & The maximum mass & $U(30,100)$  \\
$\alpha$ & Slope index of the power-law mass function & $U(-8,8)$ \\
$\delta_{\rm m}[M_{\odot}]$ & Smooth scale of the mass lower edge & $U(0,10)$ \\
$\beta_{q}$ & Slope index of the mass-ratio distribution & $U(-8,8)$ \\
\hline
\multicolumn{2}{c}{Special for \textsc{PowerLaw+Peak}}\\
$\lambda_{\rm peak}$ & Fraction in the Gaussian component & $U(0,1)$\\
$\mu_{\rm m}[M_{\odot}]$ &Mean of the Gaussian component & $U(20,50)$\\
$\sigma_{\rm m}[M_{\odot}]$ & Width of the Gaussian component & $U(1,10)$\\
\hline
\multicolumn{2}{c}{Special for \textsc{PowerLaw+Spline}}\\
$\{f_i\}_{i=2}^{11}$ & Interpolation values of perturbation for mass function & $\mathcal{N}(0,1)$ \\
\hline
\multicolumn{2}{c}{Effective-spin distribution model}\\
$\mu_{\chi,0}$ & Mean of $\chi_{\rm eff}$ distribution given $q=0.5$ & $U(-1,1)$\\
$\lg\sigma_{\chi,0}$ & Log width of $\chi_{\rm eff}$ distribution given $q=0.5$ & $U(-1.5,0.5)$\\
$a$ & Correlation between $\mu_{\chi}$ and $q$ & $U(-2.5,2.5)$ \\
$b$ & Correlation between $\log\sigma_{\chi}$ and $q$ & $U(-2,2)$ \\
\hline
\multicolumn{2}{c}{Rate evolution model} \\
$\lg (R_0[{\rm Gpc}^{-3}~{\rm yr}^{-1}])$ & Local merger rate  density & $ U(-3,3)$ \\
$z_p$ & Peak point for the rate evolution function & $U(0,4)$\\
$\gamma$ & Slope of the power-law regime before $ z_p$ & $U(-8,8)$ \\
$\kappa$ & Slope of the power-law regime after $ z_p$ & $U(-8,8)$ \\
\hline
\multicolumn{2}{c}{Transition model} \\
$m_{\rm t}[M_{\odot}]$ & The transition point in primary-mass function & $U(20,70)$ \\
$\delta_{\rm t}[M_{\odot}]$ & Smooth scale of the transition & $U(0,10)$ \\
$\mu_{\chi,2}$ & Mean of $\chi_{\rm eff}$ distribution for secondary subpopulation & $U(-1,1)$\\
$\lg\sigma_{\chi,2}$ & Log width of $\chi_{\rm eff}$ distribution secondary subpopulation  & $U(-1.5,0.5)$\\
\hline
\multicolumn{2}{c}{Mixture model} \\
$m_{\rm min,2}[M_{\odot}]$ & Minimum mass of the secondary component & $U(10,50)$ \\
$m_{\rm max,2}[M_{\odot}]$ & Maximum mass of the secondary component & $U(50,100)$\\
$\alpha_2$ & Index of $m_1$ distribution in the secondary component & $U(-8,8)$ \\
$\mu_{\chi,2}$ & Mean of $\chi_{\rm eff}$ distribution for secondary subpopulation & $U(-1,1)$\\
$\lg\sigma_{\chi,2}$ & Log width of $\chi_{\rm eff}$ distribution secondary subpopulation  & $U(-1.5,0.5)$\\
$r_2$ & fraction of the secondary component & $U(0,1)$ \\
\hline
\hline
\end{tabular}
\\
\begin{tabular}{l}
Note: $U$ and $\mathcal{N}$ are for Uniform and Normal distribution.
\end{tabular}
\end{table*} 

\section{Additional results}\label{app:results}
\begin{figure}
	\centering  
\includegraphics[width=0.8\linewidth]{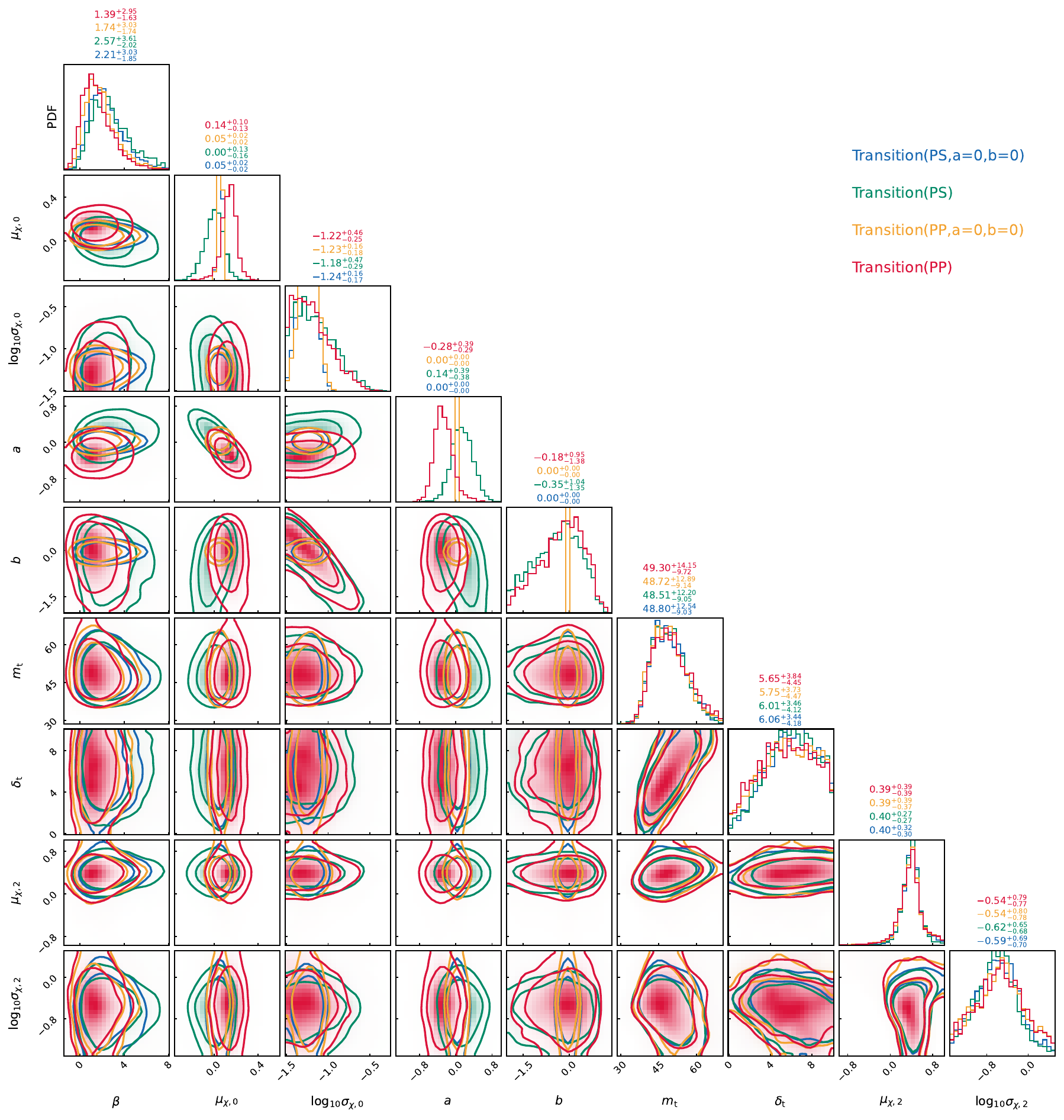}
\caption{Posteriors of the special hyperparameters for the Transition model. The contours mark the central 50\% and 90\% posterior credible regions, the values represent the median and 90\% credible intervals.}
\label{fig:mt_corner}
\end{figure}

\begin{figure}
	\centering  
\includegraphics[width=0.96\linewidth]{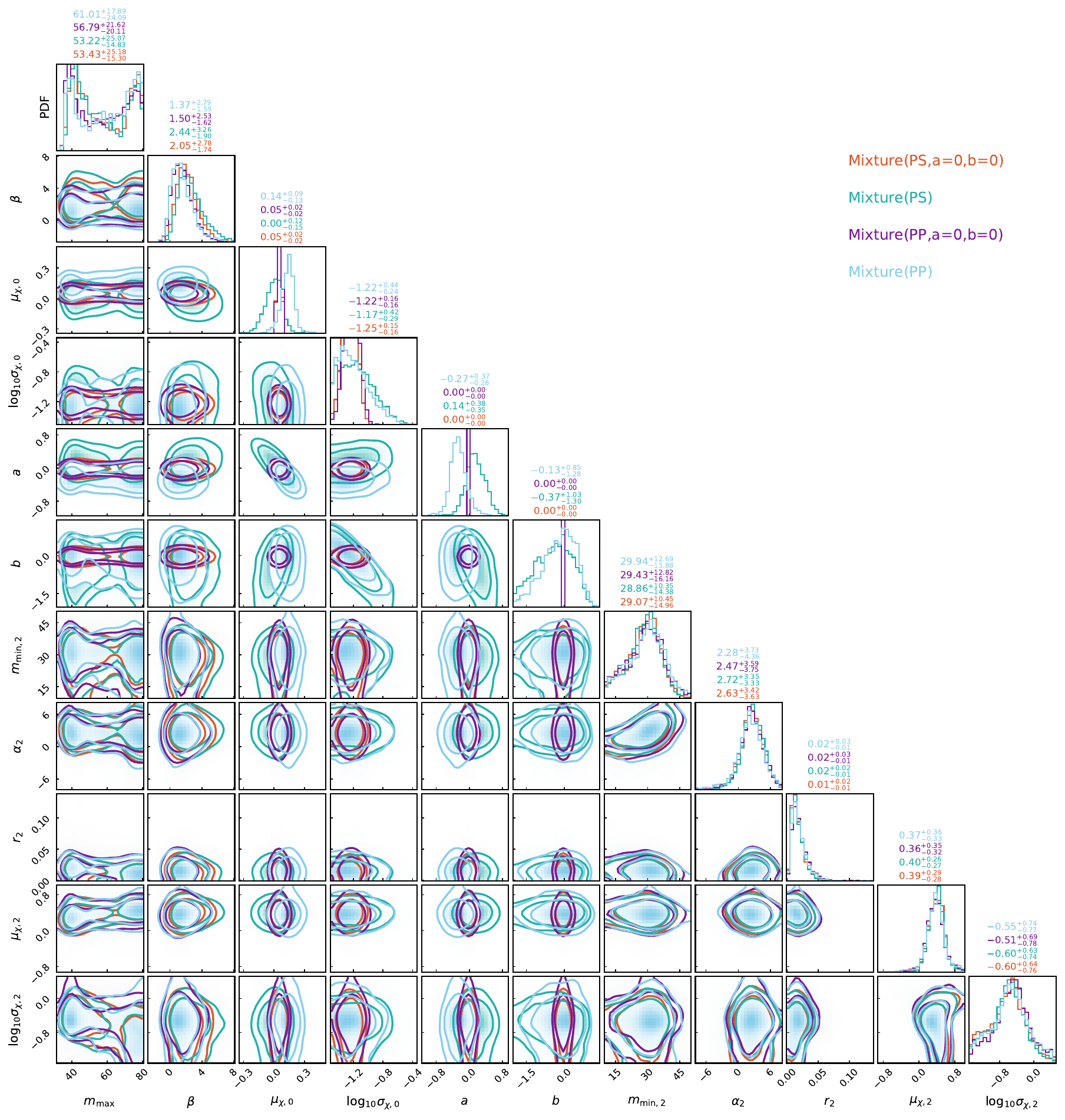}
\caption{Posteriors of the special hyperparameters for the Mixture model. The contours mark the central 50\% and 90\% posterior credible regions, the values represent the median and 90\% credible intervals.}
\label{fig:mix_corner}
\end{figure}

Figure~\ref{fig:mt_corner} and Figure~\ref{fig:mix_corner} present the full hyperparameters (except for the interpolation knots) of the Transition model, and Mixture model respectively. We find that whether we fix $a=0,b=0$ or not does not affect the transition / classification of the two subpopulations. These results are consistent with the analysis using spin-magnitude distributions in our previous work \citep{2024PhRvL.133e1401L}.
We observe that the results of the two models are well consistent with each other. The first $\chi_{\rm eff}$ distribution narrowly peaks at $\sim0.05$, while the second $\chi_{\rm eff}$ distribution peaks at $\sim0.4$, favoring a contribution from hierarchical mergers in AGN disks \citep{2019PhRvL.123r1101Y}. Note that some other formation channels, such as stable mass transfer \citep{2021A&A...647A.153B} and chemically homogeneous evolution \citep{2016MNRAS.458.2634M} can also produce the $\chi_{\rm eff}$ distributions peaking at positive values.

\begin{figure}
	\centering  
\includegraphics[width=0.4\linewidth]{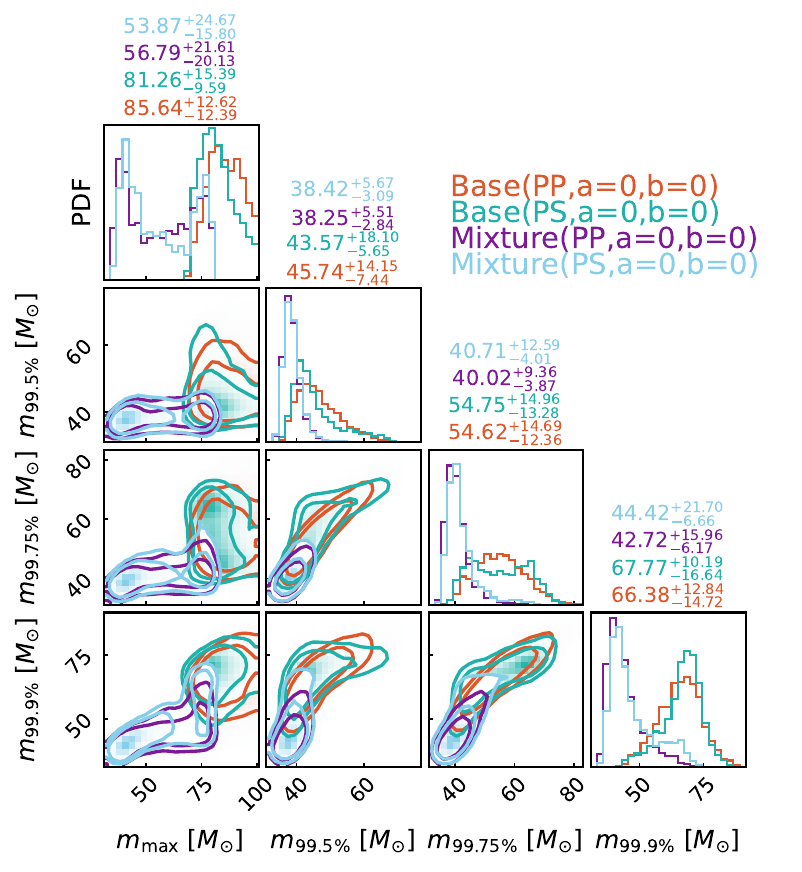}
\caption{Posteriors of the maximum mass, masses of 99.5th and 99.75th, and 99.9th percentiles of the first subpopulation for the Mixture models. The contours mark the central 50\% and 90\% posterior credible regions, the values represent the median and 90\% credible intervals.}
\label{fig:percents_corner}
\end{figure}

\begin{figure}
	\centering  
\includegraphics[width=0.96\linewidth]{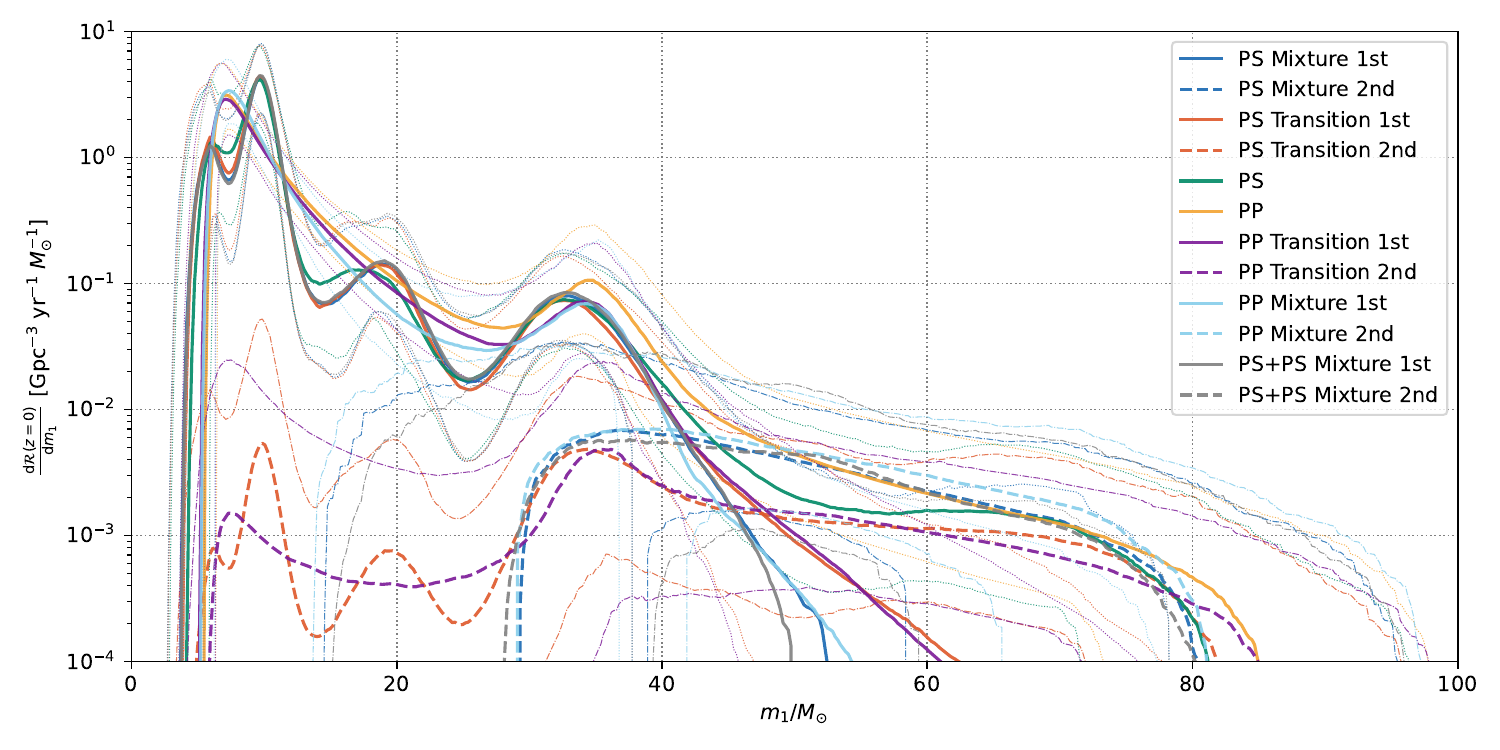}
\caption{Mass distributions inferred with various models. The thick and thin lines represent the mean values and 90\% credible intervals.}
\label{fig:mass_dist}
\end{figure}

{
For the Mixture models, the maximum mass of the first subpopulation is not well constrained but exhibits a tail extending to larger values, see Figure~\ref{fig:mix_corner}. However, the masses of 99.5th and 99.75th, and 99.9th percentiles for the first subpopulation are well measured. Especially the $m_{99\%}$ of the first subpopulation differs significantly from that of the entire population inferred from the base models, see Figure~\ref{fig:percents_corner}. 

Figure~\ref{fig:mass_dist} presents the mass distributions of the subpopulations and the entire population inferred with different models. A consistent feature across all models (Mixture and Transition) is the rapid decline of the first subpopulation's mass distribution above $\sim 40M_{\odot}$, aligning with the BH mass gap predicted by the Pair-instability Supernova \citep{2019ApJ...887...53F}.  The second subpopulation dominates the high-mass range; however, the lower boundary of its mass distribution remains uncertain.
We additionally tested a more flexible mass function for the second subpopulation in the Mixture (PS) model by applying low-mass tapering and perturbation functions to the second power-law component (labeled PS+PS). Despite this modification, the results are nearly identical to those without low-mass tapering and perturbation functions, as demonstrated in Figure~\ref{fig:mass_dist}.
}

\section{The second subpopulation}
\subsection{Is the second $\chi_{\rm eff}$ distribution symmetric?}\label{app:asym}

\begin{figure}
	\centering  
\includegraphics[width=0.4\linewidth]{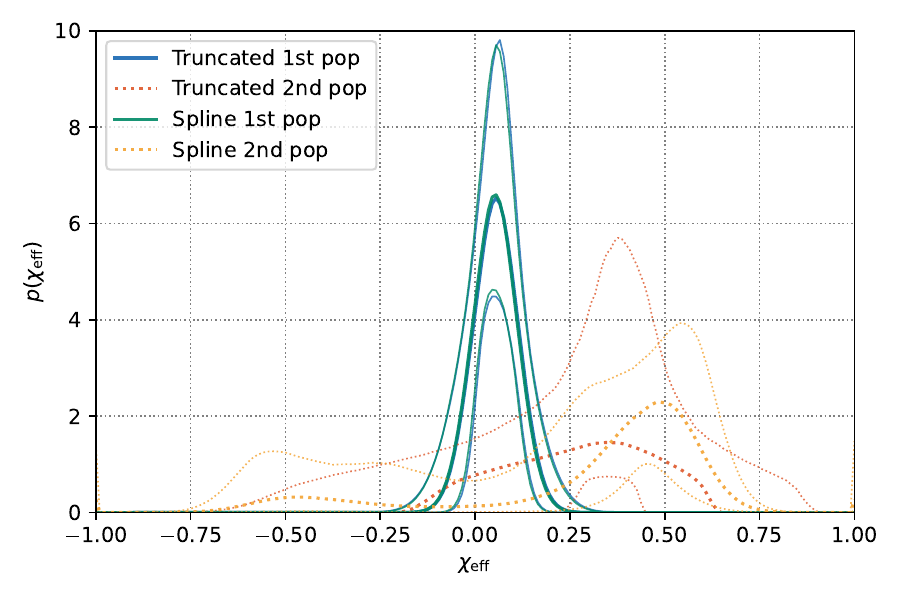}
\caption{$\chi_{\rm eff}$ distributions inferred using Transition (PS) with Truncated model and Spline model describing the second $\chi_{\rm eff}$ distribution. The thick and thin lines represent the mean values and 90\% credible intervals.}
\label{fig:tf_TG}
\end{figure}

The symmetry of $\chi_{\rm eff}$ distribution is critical for determining the formation environments of hierarchical mergers. Hierarchical mergers in star clusters always exhibit a symmetric $\chi_{\rm eff}$ distribution about zero \citep{2024ApJ...966L..16P,2022ApJ...935L..26F}. While AGN-driven hierarchical mergers tend to favor positive $\chi_{\rm eff}$ due to gas torques \citep{2018ApJ...866...66M}. 
In this section, we use two extended models to investigate the $\chi_{\rm eff}$ distribution of the second subpopulation. The accompanied mass function is the PS model. First, we infer with variable lower edge and upper edge ($\chi_{\rm min,2}$, $\chi_{\rm max,2}$) for the second $\chi_{\rm eff}$ distribution (labeled Truncated). We infer that $\chi_{\rm max,2}>0.4$ at 97\% credible level. {However, the $\chi_{\rm min,2}$ is only poorly constrained}. 
{Second, we implement a spline-based model for the second $\chi_{\rm eff}$ distribution (labeled Spline), $P_\mathcal{S}(\chi_{\rm eff}) \propto e^{f(\chi_{\rm eff})}[-1,1]$. $f (x)$ is a cubic spline function defined by 6 nodes linearly located in $[-1, 1]$. We restrict the 1-st and 6-th nodes to be -10 corresponding to $P_\mathcal{S}(\chi_{\rm eff}=1),P_\mathcal{S}(\chi_{\rm eff}=-1)\sim0$, and the priors on the amplitude of 2-nd to 5-th nodes are unit Gaussian distributions \citep{2023PhRvD.108j3009G}. 

Figure~\ref{fig:tf_TG} presents the $\chi_{\rm eff}$ distributions inferred from the Truncated model and Spline model. The $\chi_{\rm eff}$ distributions of the first subpopulation show high consistency between both models. For the second subpopulation, both models exhibit a preference for positive $\chi_{\rm eff}$ values, though negative values cannot be ruled out based on the current constraints.
}

{
\subsection{Is there $\chi_{\rm eff}-q$ correlation in the second subpopulation?}\label{app:2ab}
To test whether there is $\chi_{\rm eff}-q$ correlation in the second subpopulation, we extend the Transition model, i.e., replacing the second $\chi_{\rm eff}$ distribution $\mathcal{G}_{[-1,1]}(\chi_{\rm eff}|\mu_{\chi,2}, \sigma_{\chi,2})$ with $\mathcal{G}_{[-1,1]}(\mu_{\chi}(q;\mu_{\chi,2},a_2), \sigma_{\chi}(q;\sigma_{\chi,2},b_2))$. We find there is no evidence for the $\chi_{\rm eff}-q$ correlation in the second subpopulation, as shown in Figure~\ref{fig:2ab}. Therefore, comparing the Base models and the Transition /  Mixture models, we can conclude that the $\chi_{\rm eff}-q$ anti-correlation in the entire population is mainly attributed to the superposition of two subpopulations.

\begin{figure}
	\centering  
\includegraphics[width=0.2\linewidth]{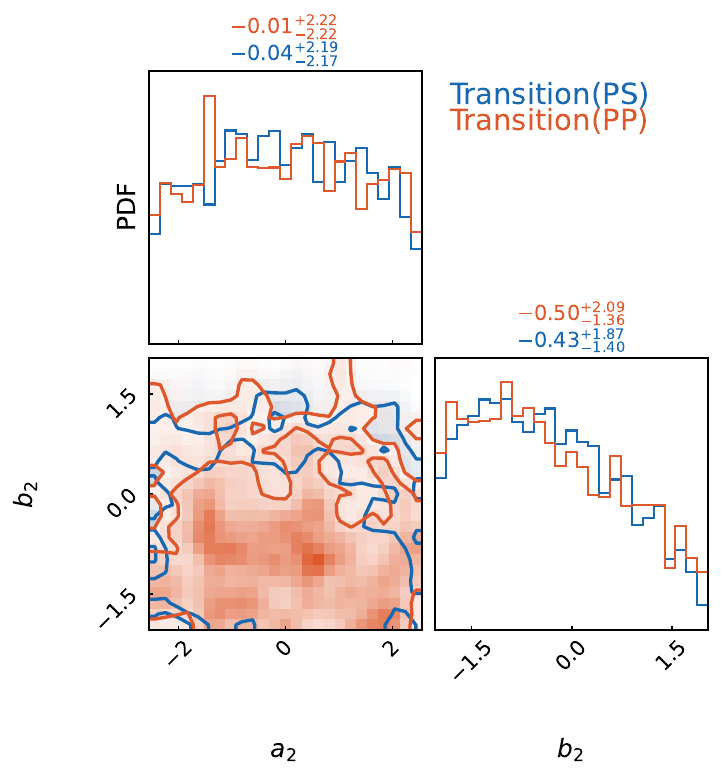}
\caption{Hyperparameters describing the $\chi_{\rm eff}-q$ correlation for the second subpopulation. The contours mark the central 50\% and 90\% posterior credible regions, the values represent the median and 90\% credible intervals.}
\label{fig:2ab}
\end{figure}
}

{
\section{Mock data studies}\label{app:mock}

To test the validity of our findings, we conduct an end-to-end injection study, including the parameter estimation and hierarchical analysis of mock signals. We generate three distinct mock populations. The first has a non-evolving $\chi_{\rm eff}$ distribution (hereafter NonEvo). The second has the same feature that inferred with the Transition model in real data (hereafter TF\&Asym). The third is similar to the second one, but for the $\chi_{\rm eff}$ of high-mass BBHs exhibiting a symmetric distribution about zero (hereafter TF\&Sym). 
Then we use the non-evolving model and Transition model to recover the features in the mock populations.

For all three mock populations, the mass distributions are generated by the PP model with $m_{\rm min}=5~M_{\odot}$, $m_{\rm max}=80~M_{\odot}$, $\delta_{\rm m}=5~M_{\odot}$, $\alpha=3.5$, $\beta=1$, $\mu_{\rm m}=35~M_{\odot}$, $\sigma_{\rm m}=3~M{\odot}$, $\lambda_{\rm peak}=0.02$, the merger rate density as a function of redshift follows Eq.~(\ref{eq:zmodel}) with $\gamma=2.7$, $z_{\rm p}=2$, and $\kappa=4$. For the NonEvo case, the $\chi_{\rm eff}$ distribution follows a Gaussian distribution $\mathcal{G}_{[-1,1]}(\mu_{\chi},\sigma_{\chi})$ with $\mu_\chi=0.06$ and $\sigma_\chi=0.1$. For the TF\&Asym case, the $\chi_{\rm eff}$ distribution exhibits mass dependence, i.e., the subpopulation with primary masses $<40~M_{\odot}$ follows a Gaussian distribution $\mathcal{G}_{[-1,1]}(\mu_{\chi,1},\sigma_{\chi,1})$ with $\mu_{\chi,1}=0.05$ and $\sigma_{\chi,1}=0.05$, and the subpopulation with primary masses $>40~M{\odot}$ follows a Gaussian distribution $\mathcal{G}_{[-1,1]}(\mu_{\chi,2},\sigma_{\chi,2})$ with $\mu_{\chi,2}=0.3$ and $\sigma_{\chi,2}=0.3$. The TF\&Sym case modifies the TF\&Asym scenario by setting $\mu_{\chi,2} = 0$ while maintaining other parameters identical..

The mock detections are sampled from the injection campaign for O3 Search Sensitivity Estimates\footnote{https://doi.org/10.5281/zenodo.7890437} with the inverse FAR $>$ 1 yr and the weight proportional to $p(\theta|{\bf \Lambda})/p_{\rm draw}(\theta)$, where $p(\theta|{\bf \Lambda})$ is probability distribution of the mock population, and $p_{\rm draw}(\theta)$ is the probability distribution from which the injection campaigns are drawn. For each case, we adopt 69 events. Following \citet{2021ApJ...922L...5C}, we set aligned spin components to $\chi_{1,z}=\chi_{2,z}=\chi_{\rm eff}$ and in-plane components to zero. Then we use IMRPhenomXAS waveform\citep{2020PhRvD.102f4001P} to generate GW signal and inject it to the noise generated by the “O3 actual” noise power spectral densities\footnote{https://dcc.ligo.org/LIGO-T2000012/public}. We perform parameter estimation on each event using BILBY \citep{2019ascl.soft01011A}, with the NESSAI nested sampler \citep{michael_j_williams_2025_14627250}.

For each case, we employ the non-evolution model and transition model to recover the mock population, i.e., Base (PP) with $a,~b=0$ and Transition (PP) with $a,~b=0$. The recovered $\chi_{\rm eff}$ distributions are presented in Figure~\ref{fig:sim} (Top). It is shown that two distinct sub-populations are successfully recognized by the transition model in the TF\&Sym and the TF\&Asym mock populations. As anticipated, the second sub-population is ambiguous for the NonEvo mock population. For both the TF\&Sym and the TF\&Asym cases, the Transition model is more favored than the Base by $\ln\mathcal{B}\sim8$. conversely, in the NonEvo case, the Transition model is less favored by $\ln\mathcal{B}\sim-1.5$.

To further assess the robustness of our analysis, we repeated the entire simulation process many times, including mock population generation, injection, recovery, and hierarchical inference. All runs produced consistent results, as illustrated in the bottom panel of Figure~\ref{fig:sim}. 

\begin{figure}
	\centering  
\includegraphics[width=0.96\linewidth]{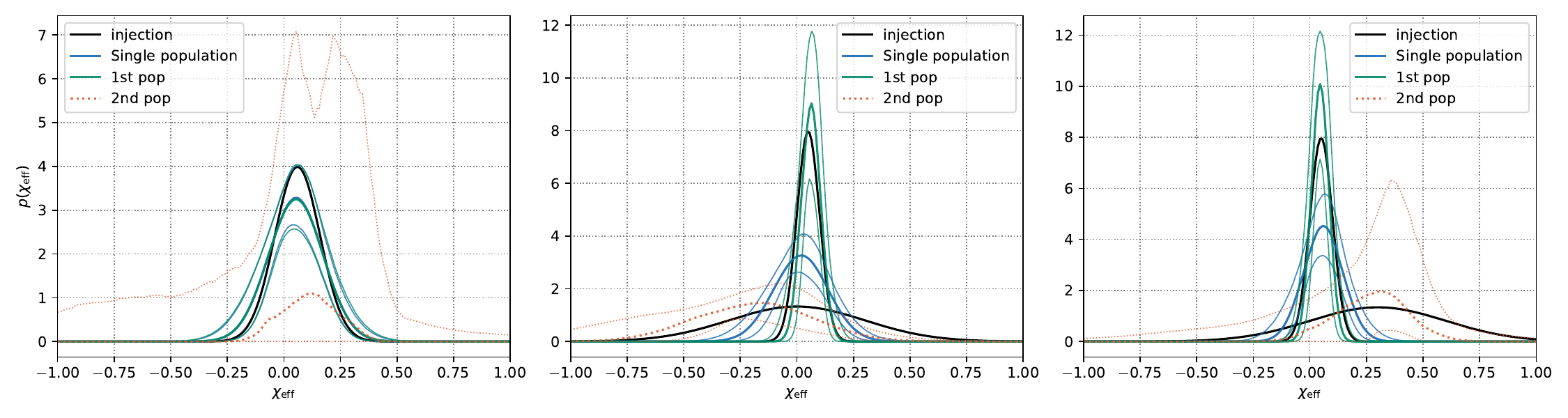}
\includegraphics[width=0.96\linewidth]{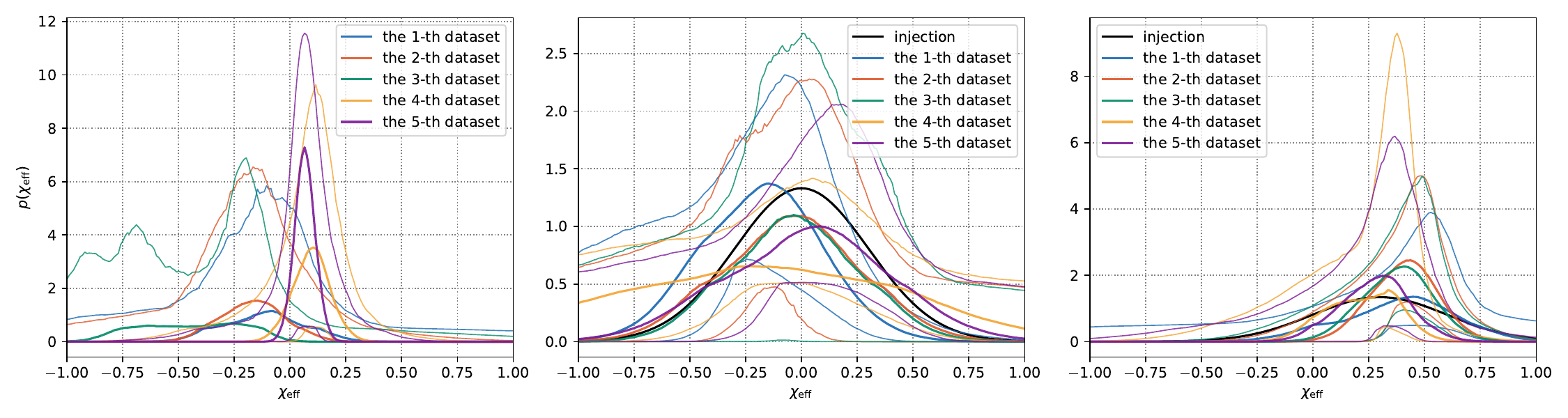}
\caption{Top: $\chi_{\rm eff}$ distributions recovered from mock population with Transition model and non-evolving model, the black lines are for the injections. Bottom: the $\chi_{\rm eff}$ distributions of the second sub-population recovered from several datasets, the black lines are for the distribution of the second sub-populations. The thick and thin lines are for the mean values and 90\% credible intervals.}
\label{fig:sim}
\end{figure}
}

\bibliography{export-bibtex}{}
\bibliographystyle{aasjournal}

\end{CJK*}
\end{document}